\documentclass[letter]{aa}     % for the letters
\usepackage{graphicx}
\usepackage{txfonts}
\usepackage{lipsum}
\usepackage{subcaption}         % necessary for continued figures, example in section 3
\usepackage{lscape}             % to rotate a single page table, example in appendix.
\usepackage{placeins}           % useful with \FloatBarrier, to keep
\usepackage{hyperref}
\usepackage{natbib}
\bibpunct{(}{)}{;}{a}{}{,} % to follow the A&A style

\newcommand{\Teff}{$T_{\rm eff}$}

\newcommand{\logl}{\ensuremath{\log(L/L_\sun)}}

\newcommand{\mj}{\ensuremath{\,M_{\rm J}}}

\newcommand{\msun}{\ensuremath{M_\sun}}

\newcommand{\teff}{$T_{\rm eff}$}
\newcommand{\rev}[1]{#1}

\begin{document}

   \title{A benchmark companion at the hydrogen-burning limit imaged in the core cluster of the Fornax-Horologium association}

   \author{Pengyu Liu\inst{1,2,3}
        \and
        Beth A. Biller\inst{2,3}
        \and
        Matthew A. Kenworthy\inst{1}
        \and
        Clémence Fontanive\inst{4}
        \and
        Ronan M. P. Kerr\inst{5}
        \and
        Tomas Stolker\inst{1}
        \and
        Christian Ginski\inst{6}
        }

   \institute{Leiden Observatory, Leiden University, PO Box 9513, 2300 RA Leiden, The Netherlands\\
              \email{pengyu.liu@ed.ac.uk}
        \and
            SUPA, Institute for Astronomy, University of Edinburgh, Royal Observatory, Blackford Hill, Edinburgh EH9 3HJ, UK
        \and
            Centre for Exoplanet Science, University of Edinburgh, Edinburgh, UK
        \and
            Trottier Institute for Research on Exoplanets, Université de Montréal, Montréal, H3C 3J7, Québec, Canada
        \and
            Dunlap Institute for Astronomy \& Astrophysics, University of Toronto, Toronto, ON M5S 3H4, Canada
        \and
             School of Natural Sciences, Centre for Astronomy, University of Galway, Galway H91 CF50, Ireland
            }

   \date{Received 14 February 2025 / Accepted 21 March 2025}

% \abstract{}{}{}{}{}
% 5 {} token are mandatory

  \abstract
   {Low-mass stellar and substellar companions are indispensable objects for verifying evolutionary models that transition from low-mass stars to planets. Their formation is likely also affected by the surrounding stellar environment.
   The Fornax-Horologium (FH) association is a recently classified young association in the solar neighbourhood with a dissolving open cluster in its centre. It has not been searched widely for substellar companions and exoplanets with direct imaging.}
   {We search for companions of stars in the FH and investigate the formation and evolution of companions during the expansion and dissolution of a star cluster.}
   {We conduct a direct-imaging survey of 49 stars in FH with VLT/SPHERE.}
   {We present HD~24121B, a companion at the hydrogen-burning limit that orbits star HD~24121 in the core cluster of the FH. The companion is located at a projected angular separation of 2.082 $\pm$ 0.004 arcsec from the central star and has contrasts of $\Delta H$2 = 5.07 $\pm$ 0.05 mag and $\Delta H$3 = 4.98 $\pm$ 0.05 mag.
   We estimate a photometric mass of 74.9 $\pm$ 7.5\,\mj{} for HD~24121B, which places it at the boundary of brown dwarfs and low-mass stars. It is a new benchmark object for evolutionary models at the hydrogen-burning boundary, especially at the age of 30 -- 40\,Myr.
   Orbital fitting with joint Gaia and SPHERE relative astrometry found a high eccentricity for HD~24121B. This suggests that it may have undergone dynamic scattering.
   Future chemical composition measurements such as high-resolution spectroscopic observations will provide more clues on its formation history.
   }
   {}


   \titlerunning{A benchmark companion HD~24121B at the hydrogen-burning limit}
   % \authorrunning{Pengyu Liu et al.}
   \maketitle

%%%%%%%%%%%%%%%%%%%%%%%%%%%%%%%%%%%%%%%%%%%%%%%%%%%%%%%%%%%%%%
\section{Introduction}
The masses of brown dwarfs are intermediate between those of low-mass stars and planets.
Brown dwarfs can serve as testbeds for substellar evolutionary models and formation mechanisms.
Direct-imaging surveys have discovered dozens of brown dwarf companions, such as PZ Tel B \citep{Biller2010}, CD~35-2722B \citep{Wahhaj2011}, HD~1160B \citep{Nielsen2012}, HR~2562B \citep{Konopacky2016}, and HD~4747B \citep{Crepp2016}.
Unlike low-mass planets, the formation of these heavy brown dwarfs via core accretion \citep{Pollack1996} in the protostellar and protoplanetary discs is challenging. They are more likely formed via direct collapse in the disc, such as by disc fragmentation \citep{Boss1997}.
Moreover, 40-50\% of the directly imaged substellar companions are more than 100\,au from away from their host stars \footnote{Date are from \url{https://exoplanet.eu/catalog/}.}.
The stellar environment may affect the formation of wide-orbit companions via gravitational instability and dynamical interactions \citep[e.g.,][]{Forgan2015,Zheng2015,Fujii2019}.
There are tentative findings on the frequency of stellar and brown dwarf companions that vary in different regions, but they remain limited to small samples \citep{Gratton2024, Gratton2025}.

Past and current direct-imaging surveys mainly focus on nearby young moving groups and OB associations \citep[e.g.,][]{Biller2013a,Nielsen2019,Desidera2021}.
In recent years, several new young clusters and associations have been established.
The $\chi^{1}$ Fornacis cluster (XFOR) was first compiled in a catalogue of open clusters with the name of Alessi 13 by \citet{Dias2002}. In this cluster, 108 member stars were identified with Gaia DR2 \citep{Zuckerman2019}.
After the release of Gaia-EDR3 \citep{Gaia2021}, its membership was extended to 164 stars, and its distance was determined to be $\sim$108\,pc \citep{Galli2021}.
This region was later expanded to define the Fornax-Horologium (FH) association of 329 candidate members with XFOR as the core cluster and a halo of stars surrounding the cluster \citep{Kerr2022}.
The FH is located in the Austral Complex, which is one of the nearest young stellar populations and contains other associations, including Tuc-Hor, Carina, and Columba \citep{Kerr2022}.

% These M dwarfs are located at the tidal radius of XFOR.
Open clusters are gravitationally bound, while young moving groups are not \citep[e.g.,][]{Gagne2024}.
Initially, XFOR was identified as one of the few young open clusters in the solar neighbourhood.
\cite{Hunt2024} reported that only 11\% of the open clusters within 250\,pc can be identified as bound open clusters by their Jacobi radius, and they classified XFOR as a moving group instead of an open cluster.
\cite{Kerr2022} reported that XFOR is likely in the virialised state and that the FH consists of a probably dissolving cluster with a large low-density halo.
Therefore, while the FH is currently not an overdense configuration, it likely formed in a much more compact configuration when we trace the motion of the stars back in time.
By contrast, there are young star associations (1--5\,Myr) in which stars formed in relatively loose environments.
While XFOR is no longer an open cluster, it therefore previously was in a gravitationally bound state and is now dissolving into a moving group.

The age of the FH as determined from isochrones, dynamics, and X-ray emissions, agrees well within 30–40\,Myr \citep{Galli2021, Kerr2022, Mamajek2016}. Known young associations and moving groups within 150\,pc either have ages younger than 30\,Myr, such as TW Haydrae \citep[$\sim$10\,Myr, ][]{Bell2015}, $\beta$ Pictoris \citep[$\sim$22\,Myr, ][]{Shkolnik2017}, Columba, and Carina \citep[$\sim$26\,Myr, ][]{Kerr2022}, or older than 40\,Myr, such as Tuc-Hor \citep[$\sim$46\,Myr, ][]{Kerr2022}, AB Doradus \citep[$\sim$150\,Myr, ][]{Bell2015}, and Argus \citep[40--50\,Myr, ][]{Zuckerman2019a}. Therefore, the FH provides a valuable opportunity for studying planetary systems at ages between 30 and 40\,Myr. The rich abundance of M dwarfs with warm debris discs in XFOR indicates that this cluster has gone through an active companion formation \citep{Zuckerman2019}.
Its young age and proximity make it well suited for direct-imaging surveys of exoplanets.

During the expansion and dissolution of the FH, there were likely frequent stellar encounters and flybys. These dynamical interactions have a significant impact on the formation and survivability of wide-orbit low-mass stellar and substellar companions.
On the one hand, frequent stellar encounters in a dense environment may trigger more gravitational instability \citep{Krumholz2012} and promote the formation of wide-orbit companions. On the other hand, they can also cause orbit instability and eject companions from their natal systems \citep{Parker2012,vanElteren2019}.

The FH occupies a unique parameter space in terms of age and cluster evolution.
We are therefore conducting a direct imaging survey of stars in the FH (PI: P. Liu).
We selected stars with $G$ brighter than 12\,mag and a renormalised unit weight error (RUWE) smaller than 1.4 to filter out possible unresolved binaries.
We checked the Tycho-2 \citep{Hog2000}, Gaia DR3 \citep{Gaia2021} and Hipparcos-Gaia \citep{Brandt2021} catalogues and excluded accelerating stars with possible stellar companions at small separations predicted by the COPAINS tool \citep{Fontanive2019}.
The sample contains 18 stars from the central core XFOR compiled in \cite{Galli2021} and 31 stars from the halo compiled in \cite{Kerr2022}.
The mass of the sample ranges between 0.8 and 2.4 \msun \citep{Kerr2022}.
The observations were taken with the Spectro-Polarimetric High-contrast Exoplanet REsearch (SPHERE) instrument \citep{Beuzit2019} in pupil-tracking mode.
We used the angular differential imaging (ADI) strategy \citep{Marois2006}.
We present the first companion imaged in this survey: an object at the hydrogen-burning limit orbiting HD~24121.

%%%%%%%%%%%%%%%%%%%%%%%%%%%%%%%%%%%%%%%%%%%%%%%%%%%%%%%%%%%%%%
\section{Observations and data analysis}
The observations of HD~24121 were conducted in the dual-band imaging mode with SPHERE/IRDIS in the $H$2 and $H$3 filters. We used the N\_ALC\_YJH\_S coronagraph to mask the central star.
The first-epoch observations of HD~24121 were taken on 4 Oct 2023 with a total time of 1.5\,h. The science time was 1.1\,h with a detector integration time (DIT) = 32\,s and a number of DIT (NDIT) = 1 for each exposure.
The airmass ranged from 1.04 to 1.08. The total field-of-view rotation was 42\fdg18.
The seeing varied between 0\farcs46 and 0\farcs70.
The second-epoch observations of HD~24121 were taken on 10 Nov 2024 with a total duration of 35 min. The observation mode was set to the same mode as for the first epoch. The science time was 0.16\,h with DIT = 32\,s and NDIT = 1.
The airmass ranged from 1.063 to 1.056. The total field-of-view rotation was 5\fdg57.
The seeing varied between 0\farcs63 and 0\farcs96.
In each epoch, we also took two separate non-coronagraphic exposures of the central star of DIT = 6\,s and NDIT = 1 with the neutral density filter ND\_1.0 at the beginning and end of observations.

%Therefore, we adopted an age of 30 $\pm$ 10\,Myr in the analysis of this work.

\begin{figure}
\centering
  \includegraphics[width=0.4\textwidth]{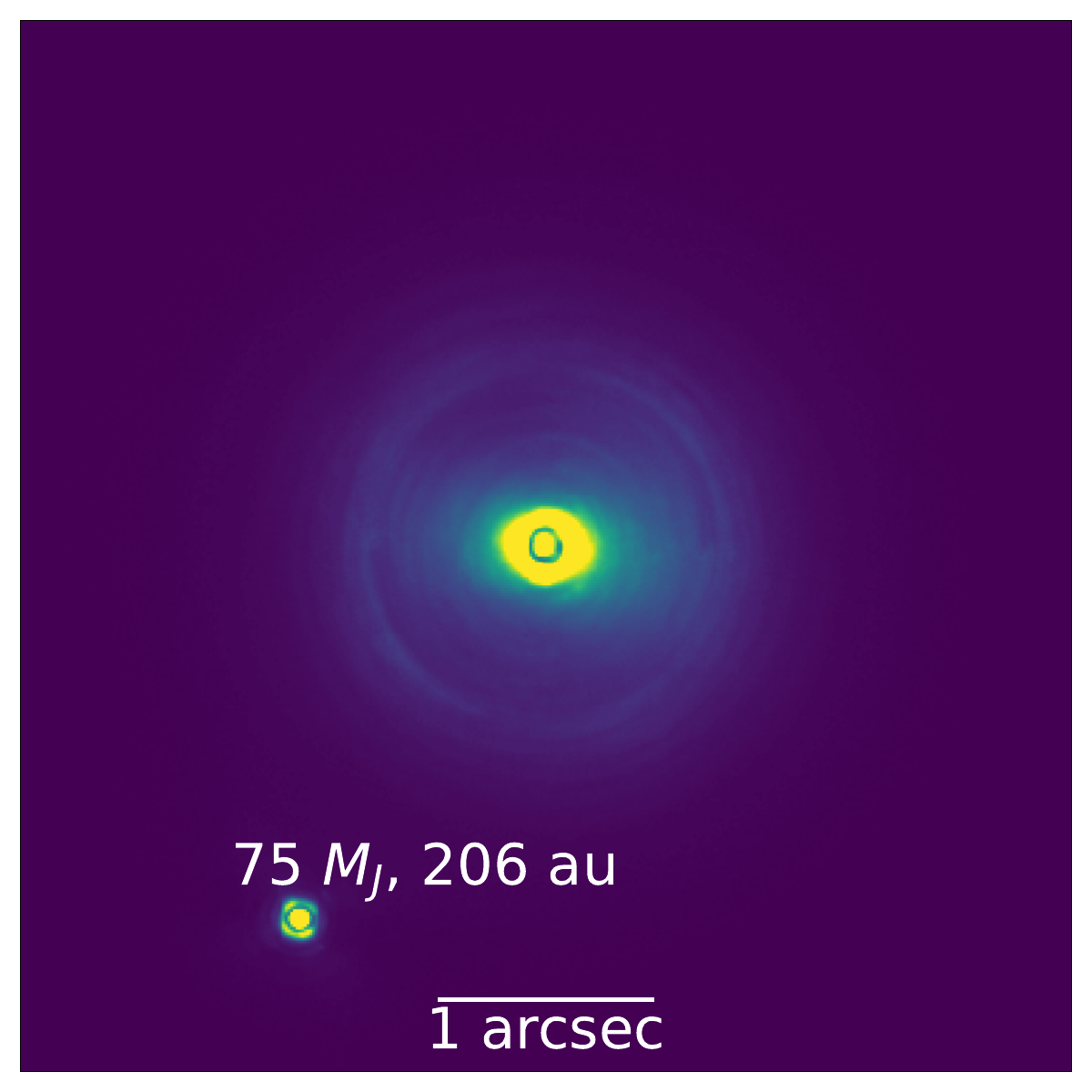}
\caption{Coronagraphic image of HD~21411 at the first epoch of SPHERE observations. The brown dwarf companion is the point source in the south-east direction. North is up, and east is to the left.
}
\label{fig:hd24121}
\end{figure}

We reduced the raw images using the ESO pipeline for SPHERE wrapped in a Python script \citep{Vigan2020}. The companion is bright enough to be seen in each preprocessed frame. It is at a distance of $\sim$2\,arcsec in the south-east direction of the central star, as shown in the de-rotated and stacked image of the first epoch in Fig.~\ref{fig:hd24121}. We also applied classical ADI to the data cube and did not detect any other sources.

\section{Results}
This companion was also detected by Gaia with the Gaia ID of Gaia~DR3~4855032616243433984 \citep{Gaia2021}. The Gaia ID of the primary star is Gaia~DR3~4855032616243433856.
They are high proper motion sources that move in similar directions in the Gaia catalogue.
Their parallax agrees within 1$\sigma$, while the difference in proper motion is larger than 2$\sigma$: -0.6 $\pm$ 0.3 mas yr$^{-1}$ in RA and -2.6 $\pm$ 0.5 mas yr$^{-1}$ in Dec, which is likely due to the orbital motion of the companion.
The primary star has a spectral type of G3 \citep{Zuckerman2019} with a mass of $\sim$1.06\,\msun \citep{Kerr2022}.
Table~\ref{tab:properties} summarises the properties of this system, updated with our SPHERE observations.

%98.3 $\pm$ 0.1

\begin{table}[h!]
\caption{Properties of the HD~24121 AB system}
\label{tab:properties}
\centering
\begin{tabular}{lcc}
\hline\hline
Property & Primary & Secondary \\
\hline
Age (Myr)\tablefootmark{a}  & 30 -- 40 & ... \\
Parallax (mas)\tablefootmark{b} &  10.17 $\pm$ 0.01 & 9.67 $\pm$ 0.53 \\
$\mu_{\alpha}cos(\delta)$ (mas yr$^{-1}$)\tablefootmark{b} & 38.72 $\pm$ 0.01 & 38.13 $\pm$ 0.34 \\
$\mu_\delta$ (mas yr$^{-1}$)\tablefootmark{b} & -2.64 $\pm$ 0.01 & -5.22 $\pm$ 0.48 \\
Sep\tablefootmark{c} (mas, 2016.0) & - & 2071.2 $\pm$ 0.4 \\
PA\tablefootmark{c} (deg, 2016.0) & - & 146.31 $\pm$ 0.01 \\
Sep\tablefootmark{d} (mas, 2023-10-04) & - & 2082 $\pm$ 4 \\
PA\tablefootmark{d} (deg, 2023-10-04) & - & 146.56 $\pm$ 0.15 \\
Sep\tablefootmark{d} (mas, 2024-11-10) & - & 2082 $\pm$ 4 \\
PA\tablefootmark{d} (deg, 2024-11-10) & - & 146.62 $\pm$ 0.15 \\
Gaia G (mag) & 9.628 $\pm$ 0.003 & 17.28 $\pm$ 0.01 \\
SPHERE $\Delta H$2 (mag)\tablefootmark{e} & - & 5.07 $\pm$ 0.05 \\
SPHERE $\Delta H$3 (mag)\tablefootmark{e} & - & 4.98 $\pm$ 0.05 \\
2MASS $J$ (mag) & 8.656 $\pm$ 0.024 & - \\
2MASS $H$ (mag) & 8.383 $\pm$ 0.046 & - \\
2MASS $K$ (mag) & 8.312 $\pm$ 0.024 & - \\
\teff(K)\tablefootmark{e} & 5819 $\pm$ 13 & 2972$^{+220}_{-198}$ \\
\logl (dex)\tablefootmark{e} & 0.012 $\pm$ 0.001 & -2.48 $\pm$ 0.06\\
Mass & 1.06 \msun \tablefootmark{f} & 74.9 $\pm$ 7.5 \mj \tablefootmark{e}\\
\hline
\end{tabular}
\tablefoot{
\tablefoottext{a}{There is no conclusive age for XFOR, but the latest age estimate is about 30\,Myr \citep{Galli2021, Kerr2022}. We adopted a conservative age range of 30 -- 40\,Myr.}
\tablefoottext{b}{Both of them have Gaia astrometry data, and we present them here.}
\tablefoottext{c}{Separation and position angle in the Gaia reference epoch 2016.0.}
\tablefoottext{d}{Separation and position angle measured in SPHERE observations.}
\tablefoottext{e}{This work.}
\tablefoottext{f}{\cite{Kerr2022}.}
}
\end{table}

\subsection{Astrometry}
\begin{figure}
\centering
  \includegraphics[width=0.5\textwidth]{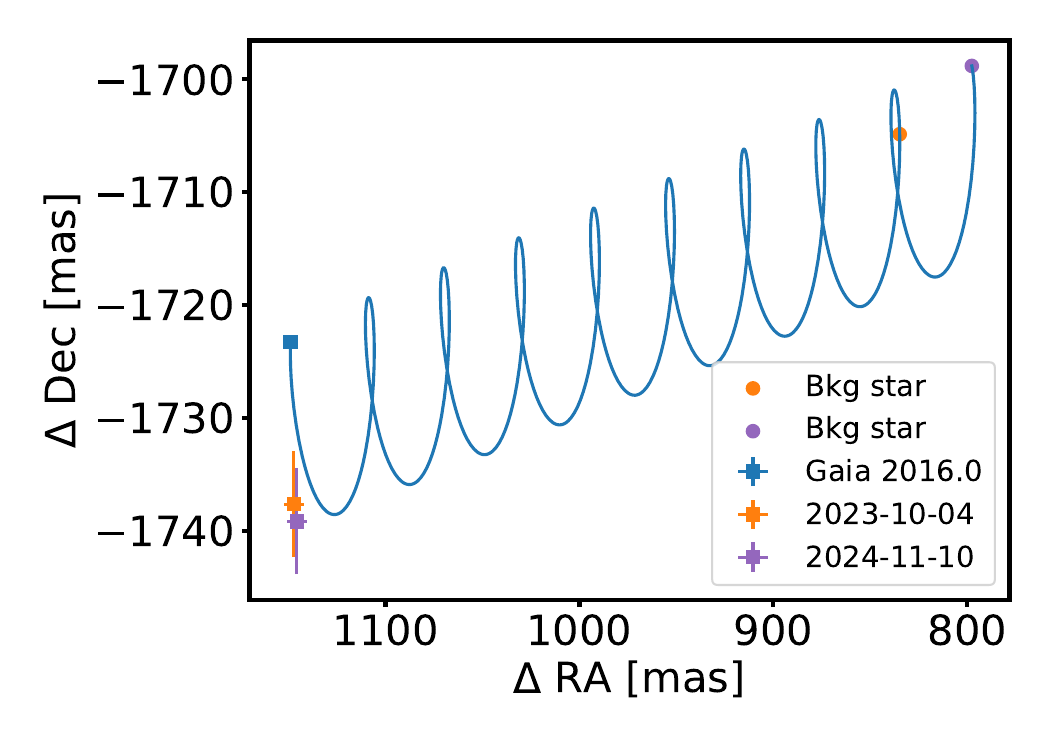}
\caption{Relative positions of HD~24121B to the central star. The blue trajectory shows the relative position of a static background star over time, starting from the Gaia reference epoch 2016.0. The orange and purple dots on the track indicate the expected positions at epochs 2023-10-04 and 2024-11-10 if HD~24121B was a static background star.
}
\label{fig:compm}
\end{figure}

We fitted a 2D Gaussian to each individual SPHERE frame and took the mean value as the measured position and the standard deviation as the fitting uncertainty. We also included the uncertainty from the pixel scale, centring, distortion, true north angle, and pupil offset following the astrometry calibration provided in the latest SPHERE User manual\footnote{\url{https://www.eso.org/sci/facilities/paranal/instruments/sphere/doc.html}} and by \cite{Maire2021}.
The results are presented in Table~\ref{tab:properties}.
We present the relative astrometry measured with SPHERE in 2023 and 2024 in Fig.~\ref{fig:compm}. We also include the relative positions of HD~24121B to HD~24121 in the Gaia reference epoch 2016.0.
It is comoving with the central star between 2023 and 2024.
The position difference between the Gaia and SPHERE epochs is likely due to the orbital motion of HD~24121B over 8 years. The projected separation is about 200\,au given the measured angular separation of $\sim$2\arcsec at a distance of 98\,pc.

\subsection{Photometry}
We performed aperture photometry of HD~24121B in the de-rotated and stacked image and subtracted the local sky background.
We calculated the signal-to-noise ratio (S/N) of HD~24121B by performing photometry on all apertures of the same size at the same separation as the source, but at different position angles, except for the aperture on the companion. Then, we took the three $\sigma$ clipped standard deviation as the noise and the average as the local sky background.
We also performed aperture photometry on the non-coronagraphic image with the same aperture radius to measure the star flux. Then, we converted the flux contrast into magnitude contrast. The magnitude contrasts measured in the first epoch are $\Delta H$2 = 5.09 $\pm$ 0.11\,mag and $\Delta H$3 = 4.99 $\pm$ 0.11\,mag. The magnitude contrasts measured in the second epoch are $\Delta H$2 = 5.07 $\pm$ 0.05\,mag and $\Delta H$3 = 4.98 $\pm$ 0.05\,mag. They agree with each other within 1$\sigma$ without discernible variability.

To estimate the effective temperature of HD~24121B, we fitted the spectral energy distribution (SED) using \textsc{species} \citep{Stolker2020b}. Data are available in three photometric bands: Gaia $G$, and SPHERE $H$2 and $H$3. We measured the contrasts of the companion to the central star in $H$2 and $H$3. To convert the SPHERE band contrasts into the apparent magnitudes of the companion, we first performed synthetic photometry on the central star. We fitted the SED of the host star using its GALEX, Tycho, SDSS, Gaia, 2MASS, and WISE photometric data with the BT-Settl-CIFIST models \citep{Allard2013}. We fitted \teff = 5819\,K, similar to 5805\,K reported in Gaia. The SED fitting results of the primary star can be found in Appendix~\ref{app:sed}. Then, we measured the primary magnitude in $H$2 and $H$3 by applying synthetic photometry on the best-fit model spectrum. Adding the contrasts measured from direct imaging, we obtained the apparent magnitude of HD~24121B in $H$2 and $H$3.

Then, we used two evolutionary models for warm brown dwarfs to fit its SED: BT-Settl \citep{Allard2003} and AMES-DUSTY \citep{Allard2001, Chabrier2000}.
We included error inflation factors in the fitting because the uncertainty from the synthetic photometry is underestimated.
We derived similar results for the two models. BT-Settl has slightly smaller residuals in the SED fitting, and therefore, we present its best-fit spectra in Fig.~\ref{fig:comsed} and the posterior distribution in Appendix~\ref{app:sed}. We fitted \Teff = 2972$^{+220}_{-198}$\,K and \logl = -2.48 $\pm$ 0.06\,dex for the companion.

\begin{figure}
\centering
  \includegraphics[width=0.5\textwidth]{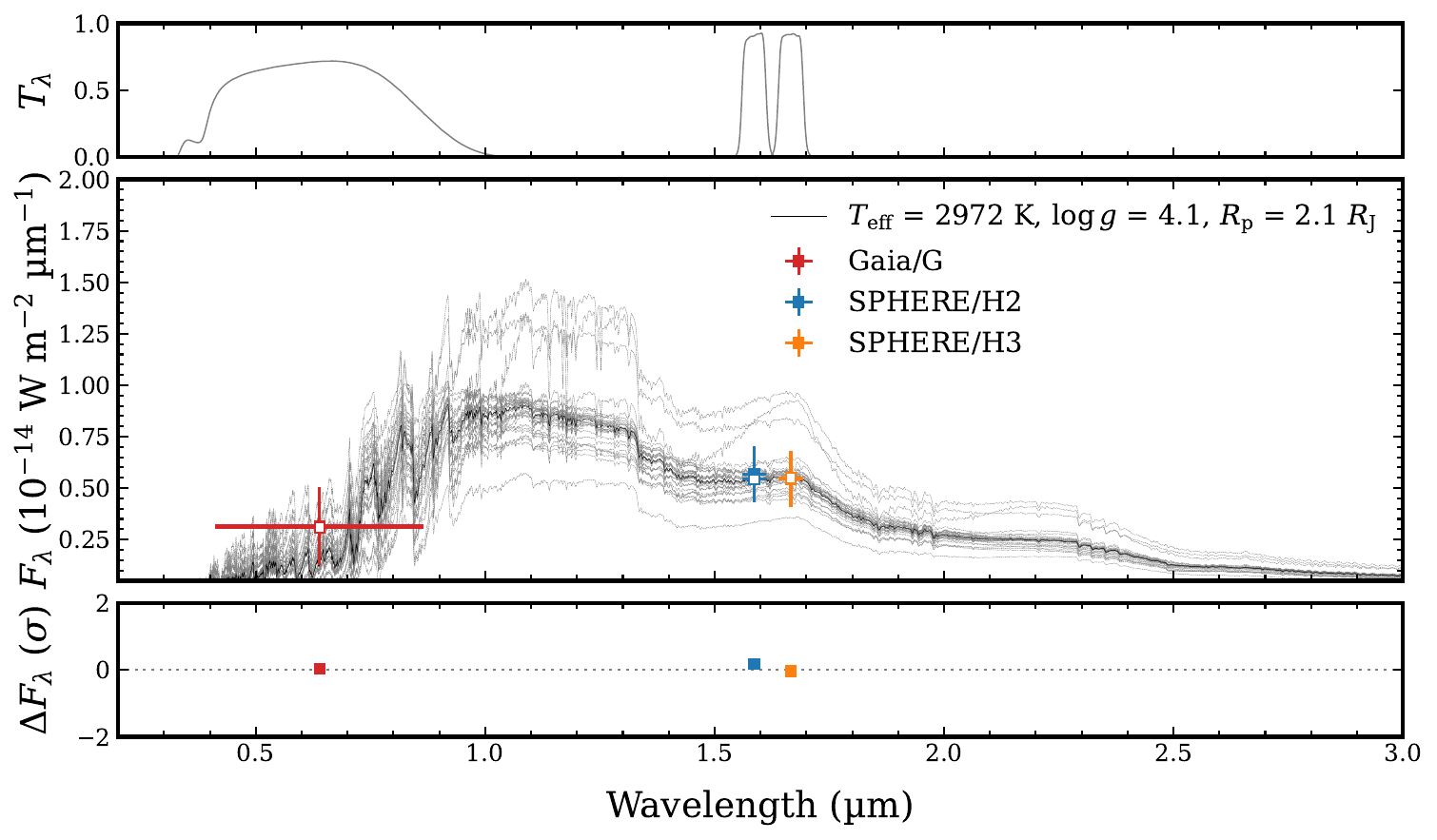}
\caption{SED fitting of the companion, combining the Gaia $G$, and SPHERE $H$2 and $H$3 photometric bands using BT-Settl atmospheric models. \rev{The filled squares are the measured photometric data with error bars, and the open squares are the modelled photometric data. The grey lines are 30 spectra randomly drawn from the posterior distributions. The upper panel shows the transmission of photometric bands, and the lower panel shows the residuals in $\sigma$.}
}
\label{fig:comsed}
\end{figure}

\subsection{Orbital fitting}
To find plausible orbits for this system, we used \textsc{orbitize} \citep{Blunt2020} to fit the relative astrometry of HD~24121B and HD~24211 by combining the Gaia reference epoch of 2016.0 and the two SPHERE epochs. We adopted a parallax of 10.17 $\pm$ 0.01\,mas for the primary star from Gaia DR3, and we assumed a total mass of 1.13 $\pm$ 0.1\,\msun. The fitted orbits are shown in Fig.~\ref{fig:orbits}.
The posterior distributions of orbital parameters are presented in Appendix~\ref{app:vel}.
We find a semi-major axis of $\sim$205\,au and a well-constrained inclination of 70\fdg4$^{+9.6}_{-19.1}$. The orbital period is $\sim$2745 year. We find a high eccentricity for HD~24121B, with a median of 0.6 and a peak of 0.9, according to its posterior distribution.

We calculated the fitted orbital velocity of HD~24121B in the Gaia reference epoch and derived the velocity in RA (v\_ra) = -0.3 $\pm$ 0.4\,mas yr$^{-1}$ and the velocity in Dec (v\_dec) = -1.8 $\pm$ 0.4\,mas yr$^{-1}$. The posterior distribution is presented in Appendix~\ref{app:vel}. The fitted orbital velocity matches the Gaia measured proper motion difference within 1--2$\sigma$. Therefore, the two sources are very likely in a gravitationally bound system.

\begin{figure}
\centering
  \includegraphics[width=0.5\textwidth]{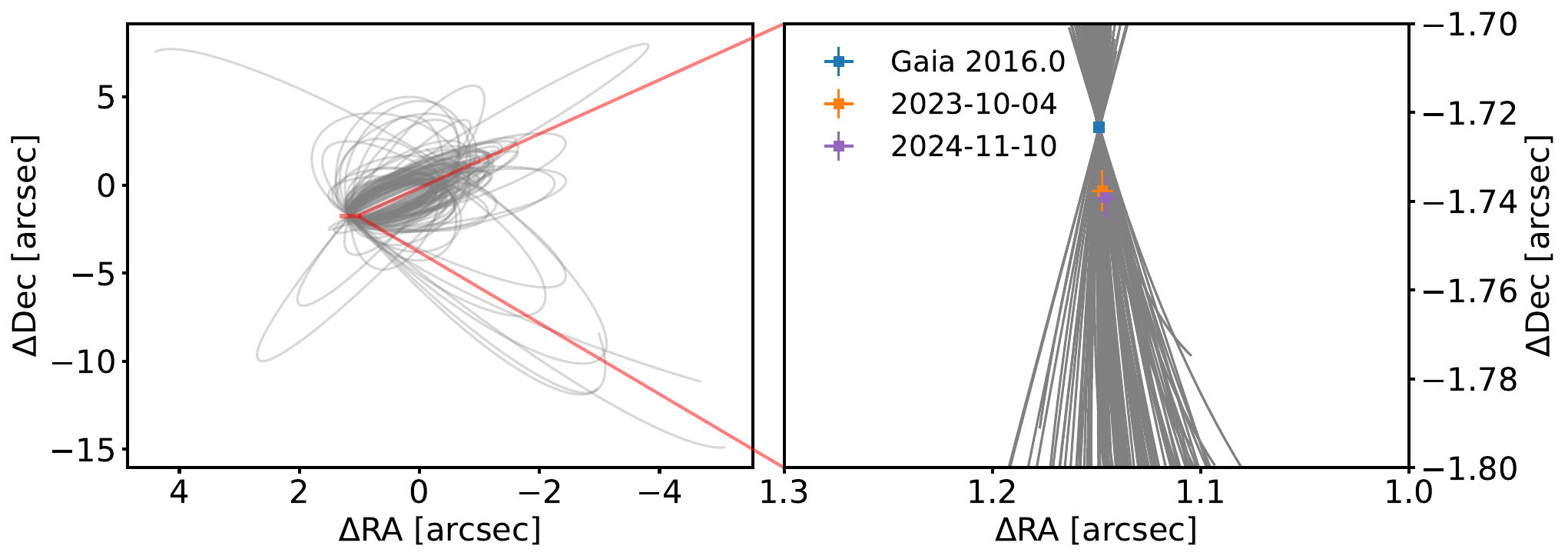}
\caption{Fitted orbits of HD~24121B. We show 100 random orbits drawn from the posterior of the fit.
}
\label{fig:orbits}
\end{figure}

\subsection{Companion mass estimation}
\begin{figure}
\centering
  \includegraphics[width=0.5\textwidth]{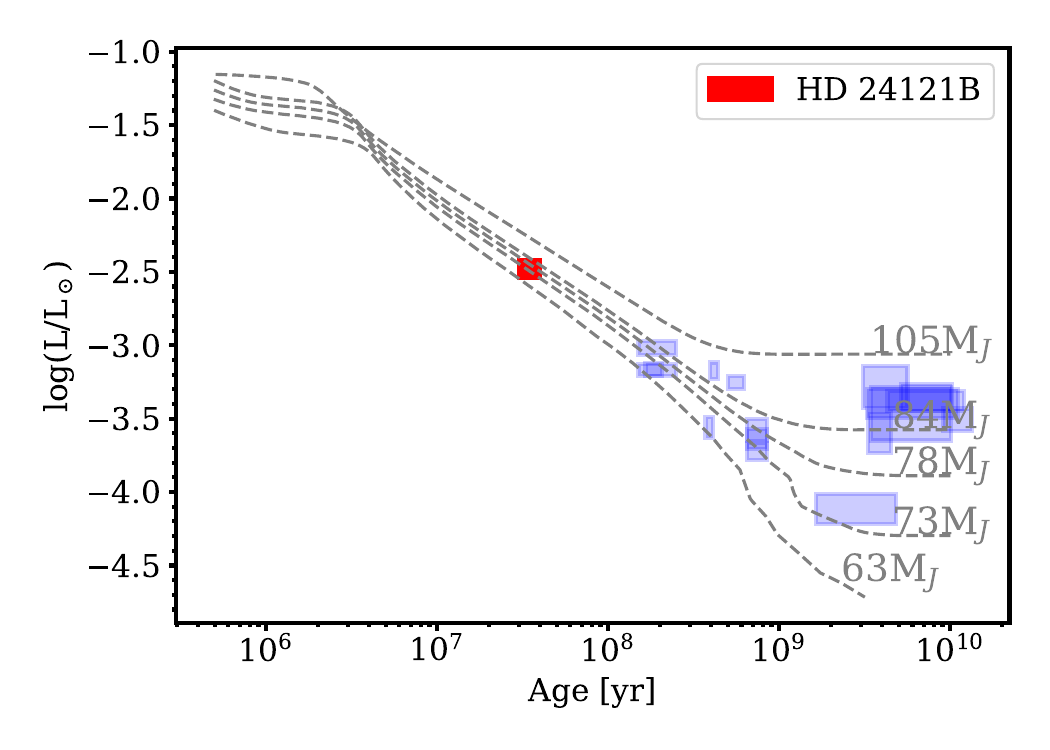}
\caption{Evolutionary tracks of low-mass stars and brown dwarfs from the BCAH15 models. Hydrogen-burning stars reach stable luminosities after a certain contraction time, while brown dwarfs continue to cool.
The red square indicates the luminosity (1$\sigma$) and age (30 -- 40\,Myr) range of HD~24121B.
The blue squares show the companions at the hydrogen-burning limit from the Gaia ultracool dwarf companion catalogue, which includes objects within 100\,pc \citep{Baig2024}.
HD~24121B is located at the hydrogen-burning limit between low-mass stars and brown dwarfs, and it is much younger than the other companions.
}
\label{fig:lumage}
\end{figure}

The dynamical mass of the companion cannot be constrained in the orbital fitting due to the small orbit coverage.
There is no evidence of Gaia acceleration in either source: Their RUWE is smaller than 1. There is no proper motion anomaly between the Gaia and Hipparcos catalogue \citep{Brandt2021}.
Therefore, we estimated the mass from photometry using the rejection-sampling method \citep[e.g.,][]{Dupuy2017}. With a bolometric luminosity of -2.48 $\pm$ 0.06 dex from the SED fitting and an age of 32.4 $\pm$ 1.3\,Myr \citep{Kerr2022}, the derived mass is 72.2 $\pm$ 6.4\,\mj{} using the BT-Settl evolutionary models.
With a broader uniform age distribution between 30 and 40 Myr, the derived mass is 74.9 $\pm$ 7.5\,\mj{}.
We obtained consistent results with the BCAH15 model \citep{Baraffe2015}.
As demonstrated in Fig.~\ref{fig:lumage}, HD~24121 straddles the region between brown dwarfs and low-mass hydrogen-burning stars in the evolutionary tracks.

%%%%%%%%%%%%%%%%%%%%%%%%%%%%%%%%%%%%%%%%%%%%%%%%%%%%%%%%%%%%%%
\section{Discussions and conclusions}
We presented a companion of HD~24121, the object HD~24121B.
With a mass of 74.9 $\pm$ 7.5\,\mj{}, HD~24121B can serve as a benchmark object in characterising the evolutionary models of low-mass stars and brown dwarfs at the hydrogen-burning boundary.
There are not many directly imaged low-mass stellar and substellar companions at the age of 30\,Myr.
Almost all of the benchmark brown dwarf companions at the hydrogen-burning limit were found in relatively old systems such as HD~72946B \citep[72.4 $\pm$ 1.6\,\mj{}, 0.9 -- 3\,Gyr,][]{Maire2020}, HR~7672B \citep[$\sim$68.7\,\mj{}, 2.5\,Gyr,][]{Crepp2012}, and HD~4747B \citep[70 $\pm$ 1.6 \mj{}, 2.3 $\pm$ 1.4\,Gyr][]{peretti2019}, and those compiled in the Gaia ultracool dwarf companion sample \citep{Baig2024}.
Most of the directly imaged companions in the literature that are younger than 30\,Myr have masses lower than 70\,\mj{}, which is likely due to a selection bias.
Nevertheless, HD~24121B provides a unique age range for studying evolutionary models at the hydrogen-burning limit.
At the age of 30\,Myr with a mass of 75\,\mj{}, HD~24121B is about to start lithium burning according to the BT-Settl and BCAH15 models.
A future spectroscopic characterisation of HD~24121B can measure its lithium abundance.

The formation of HD~24121B is also an intriguing question. It is challenging to form such a heavy companion at 200\,au by core accretion or even pebble accretion \citep{Chambers2021,Savvidou2023}.
Disc fragmentation triggered by gravitational instability and binary-like cloud fragmentation are probably the most favourable formation mechanisms here. They can produce brown dwarfs and low-mass hydrogen-burning stellar companions around Sun-like stars \citep{Stamatellos2009}.
Moreover, subsequent dynamical effects such as fragment-fragment scattering are efficient powers to modify the orbital parameters of companions \citep{Hall2017,Forgan2018}. The fragment-fragment scattering can populate the companions to a broader semi-major axis range such as over $10^2 - 10^3$ au \citep{Forgan2018}. Dynamical evolution including fragment-fragment scattering and multi-object scattering also result in a relatively higher eccentricity for wide-orbit companions or even planet ejection and recapturing \citep{Forgan2015}, which agrees with the high eccentricity we derived for HD~24121B.
\cite{Hwang2022} reported that wide stellar binaries have a uniform eccentricity distribution for separations within 200\,au. A similar trend has been found for brown dwarfs within 100\,au: They have a broad peak at an eccentricity of 0.6--0.9 \citep{Bowler2020}.
Therefore, HD~24121B is likely formed by disc or cloud fragmentation, followed by a dynamical evolution, before it reached its current orbit.

The possible dynamical scattering might result from interactions with another companion in the same system.
To estimate the upper mass limit of the potential another companion, we calculated the 5$\sigma$ contrast curve of the dataset of the epoch 4 Oct 2023 and reached a sensitivity of 3\,\mj{} beyond $\sim$1\farcs5.
Therefore, this companion is probably smaller than 3\,\mj{} or at a smaller separation with a higher mass if it has survived in the same system.
Because HD~24121 is located in the core cluster of the FH, dynamical scattering between HD~24121 and other stars may also have been possible.
The core cluster of the FH is a dissolving open cluster, and its core region is about one order of magnitude denser than the local disc stellar density \citep{Zuckerman2019}. When we trace the motion of stars back, the chances of stellar encounters and flybys are probably much higher than in the current stage.

A composition and abundance comparison of HD~24121 and HD~24121B can give more clues about the formation of HD~24121B, such as the C/O ratio, which can serve as an indicator of its formation location in the disc and evidence of gas accretion \citep[e.g.,][]{Oberg2011}.
Recently, the carbon isotopologue ratio was proposed as a probe to distinguish low-mass planet and brown dwarf formation pathways \citep{Zhangyp2021a, Zhangyp2021b, deRegt2024, Gonzalez2025}.
Although the central star HD~24121 may be too hot for us to measure its carbon isotopologue ratio, high-resolution spectroscopic observations of HD~24121B in the near-infrared band can constrain its isotopologue ratio, which can then be compared with that of other brown dwarfs.
Spectroscopic observations can also provide the spin-rate measurement of HD~24121B from broadened spectral lines, which will provide indications about its angular momentum evolution \citep{Bouvier2014,Schwarz2016}.
We therefore recommend follow-up spectroscopic observations to characterise HD~24121B. It can serve as a benchmark object for studying the evolutionary models and formation mechanisms of low-mass stellar and substellar companions.

% ESO~207-61 is a free-floating brown dwarf in the Haydes \citep{Ruiz1991}.

%%%%%%%%%%%%%%%%%%%%%%%%%%%%%%%%%%%%%%%%%%%%%%%%%%%%%%%%%%%%%%
\begin{acknowledgements}
We are grateful to W. K. M. Rice for insightful discussions that contributed to the development of this work.
RK acknowledges the support from the Dunlap Institute, which is funded through an endowment established by the David Dunlap family and the University of Toronto.
Based on observations collected at the European Organisation for Astronomical Research in the Southern Hemisphere under ESO programmes 112.25SC.001 and 114.2776.001.
This research made use of \textsc{Astropy}, a community-developed core \textsc{Python} package for Astronomy \citep{Astropy2013,Astropy2018}, \textsc{NumPy} \citep{Numpy2020} and \textsc{Matplotlib}, a Python library for publication quality graphics \citep{Matplotlib2007}.
\end{acknowledgements}

% Please note that we have included the references below in
% order to compile the document, but we ask you to:

\bibliographystyle{aa}
\bibliography{main}

\begin{thebibliography}{66}
\expandafter\ifx\csname natexlab\endcsname\relax\def\natexlab#1{#1}\fi

\bibitem[{{Allard} {et~al.}(2003){Allard}, {Guillot}, {Ludwig}, {Hauschildt}, {Schweitzer}, {Alexander}, \& {Ferguson}}]{Allard2003}
{Allard}, F., {Guillot}, T., {Ludwig}, H.-G., {et~al.} 2003, in IAU Symposium, Vol. 211, Brown Dwarfs, ed. E.~{Mart{\'\i}n}, 325

\bibitem[{{Allard} {et~al.}(2001){Allard}, {Hauschildt}, {Alexander}, {Tamanai}, \& {Schweitzer}}]{Allard2001}
{Allard}, F., {Hauschildt}, P.~H., {Alexander}, D.~R., {Tamanai}, A., \& {Schweitzer}, A. 2001, \apj, 556, 357

\bibitem[{{Allard} {et~al.}(2013){Allard}, {Homeier}, {Freytag}, {Schaffenberger}, \& {Rajpurohit}}]{Allard2013}
{Allard}, F., {Homeier}, D., {Freytag}, B., {Schaffenberger}, W., \& {Rajpurohit}, A.~S. 2013, Memorie della Societa Astronomica Italiana Supplementi, 24, 128

\bibitem[{{Astropy Collaboration} {et~al.}(2018){Astropy Collaboration}, {Price-Whelan}, {Sip{\H o}cz}, {G{\"u}nther}, {Lim}, {Crawford}, {Conseil}, {Shupe}, {Craig}, {Dencheva}, {Ginsburg}, {VanderPlas}, {Bradley}, {P{\'e}rez-Su{\'a}rez}, {de Val-Borro}, {Aldcroft}, {Cruz}, {Robitaille}, {Tollerud}, {Ardelean}, {Babej}, {Bach}, {Bachetti}, {Bakanov}, {Bamford}, {Barentsen}, {Barmby}, {Baumbach}, {Berry}, {Biscani}, {Boquien}, {Bostroem}, {Bouma}, {Brammer}, {Bray}, {Breytenbach}, {Buddelmeijer}, {Burke}, {Calderone}, {Cano Rodr{\'{\i}}guez}, {Cara}, {Cardoso}, {Cheedella}, {Copin}, {Corrales}, {Crichton}, {D'Avella}, {Deil}, {Depagne}, {Dietrich}, {Donath}, {Droettboom}, {Earl}, {Erben}, {Fabbro}, {Ferreira}, {Finethy}, {Fox}, {Garrison}, {Gibbons}, {Goldstein}, {Gommers}, {Greco}, {Greenfield}, {Groener}, {Grollier}, {Hagen}, {Hirst}, {Homeier}, {Horton}, {Hosseinzadeh}, {Hu}, {Hunkeler}, {Ivezi{\'c}}, {Jain}, {Jenness}, {Kanarek}, {Kendrew}, {Kern}, {Kerzendorf}, {Khvalko}, {King}, {Kirkby}, {Kulkarni},
  {Kumar}, {Lee}, {Lenz}, {Littlefair}, {Ma}, {Macleod}, {Mastropietro}, {McCully}, {Montagnac}, {Morris}, {Mueller}, {Mumford}, {Muna}, {Murphy}, {Nelson}, {Nguyen}, {Ninan}, {N{\"o}the}, {Ogaz}, {Oh}, {Parejko}, {Parley}, {Pascual}, {Patil}, {Patil}, {Plunkett}, {Prochaska}, {Rastogi}, {Reddy Janga}, {Sabater}, {Sakurikar}, {Seifert}, {Sherbert}, {Sherwood-Taylor}, {Shih}, {Sick}, {Silbiger}, {Singanamalla}, {Singer}, {Sladen}, {Sooley}, {Sornarajah}, {Streicher}, {Teuben}, {Thomas}, {Tremblay}, {Turner}, {Terr{\'o}n}, {van Kerkwijk}, {de la Vega}, {Watkins}, {Weaver}, {Whitmore}, {Woillez}, {Zabalza}, \& {Astropy Contributors}}]{Astropy2018}
{Astropy Collaboration}, {Price-Whelan}, A.~M., {Sip{\H o}cz}, B.~M., {et~al.} 2018, \aj, 156, 123

\bibitem[{{Astropy Collaboration} {et~al.}(2013){Astropy Collaboration}, {Robitaille}, {Tollerud}, {Greenfield}, {Droettboom}, {Bray}, {Aldcroft}, {Davis}, {Ginsburg}, {Price-Whelan}, {Kerzendorf}, {Conley}, {Crighton}, {Barbary}, {Muna}, {Ferguson}, {Grollier}, {Parikh}, {Nair}, {Unther}, {Deil}, {Woillez}, {Conseil}, {Kramer}, {Turner}, {Singer}, {Fox}, {Weaver}, {Zabalza}, {Edwards}, {Azalee Bostroem}, {Burke}, {Casey}, {Crawford}, {Dencheva}, {Ely}, {Jenness}, {Labrie}, {Lim}, {Pierfederici}, {Pontzen}, {Ptak}, {Refsdal}, {Servillat}, \& {Streicher}}]{Astropy2013}
{Astropy Collaboration}, {Robitaille}, T.~P., {Tollerud}, E.~J., {et~al.} 2013, \aap, 558, A33

\bibitem[{{Baig} {et~al.}(2024){Baig}, {Smart}, {Jones}, {Gagn{\'e}}, {Pinfield}, {Cheng}, \& {Moranta}}]{Baig2024}
{Baig}, S., {Smart}, R.~L., {Jones}, H. R.~A., {et~al.} 2024, \mnras, 533, 3784

\bibitem[{{Baraffe} {et~al.}(2015){Baraffe}, {Homeier}, {Allard}, \& {Chabrier}}]{Baraffe2015}
{Baraffe}, I., {Homeier}, D., {Allard}, F., \& {Chabrier}, G. 2015, \aap, 577, A42

\bibitem[{{Bell} {et~al.}(2015){Bell}, {Mamajek}, \& {Naylor}}]{Bell2015}
{Bell}, C. P.~M., {Mamajek}, E.~E., \& {Naylor}, T. 2015, \mnras, 454, 593

\bibitem[{{Beuzit} {et~al.}(2019){Beuzit}, {Vigan}, {Mouillet}, {Dohlen}, {Gratton}, {Boccaletti}, {Sauvage}, {Schmid}, {Langlois}, {Petit}, {Baruffolo}, {Feldt}, {Milli}, {Wahhaj}, {Abe}, {Anselmi}, {Antichi}, {Barette}, {Baudrand}, {Baudoz}, {Bazzon}, {Bernardi}, {Blanchard}, {Brast}, {Bruno}, {Buey}, {Carbillet}, {Carle}, {Cascone}, {Chapron}, {Charton}, {Chauvin}, {Claudi}, {Costille}, {De Caprio}, {de Boer}, {Delboulb{\'e}}, {Desidera}, {Dominik}, {Downing}, {Dupuis}, {Fabron}, {Fantinel}, {Farisato}, {Feautrier}, {Fedrigo}, {Fusco}, {Gigan}, {Ginski}, {Girard}, {Giro}, {Gisler}, {Gluck}, {Gry}, {Henning}, {Hubin}, {Hugot}, {Incorvaia}, {Jaquet}, {Kasper}, {Lagadec}, {Lagrange}, {Le Coroller}, {Le Mignant}, {Le Ruyet}, {Lessio}, {Lizon}, {Llored}, {Lundin}, {Madec}, {Magnard}, {Marteaud}, {Martinez}, {Maurel}, {M{\'e}nard}, {Mesa}, {M{\"o}ller-Nilsson}, {Moulin}, {Moutou}, {Orign{\'e}}, {Parisot}, {Pavlov}, {Perret}, {Pragt}, {Puget}, {Rabou}, {Ramos}, {Reess}, {Rigal}, {Rochat}, {Roelfsema}, {Rousset},
  {Roux}, {Saisse}, {Salasnich}, {Santambrogio}, {Scuderi}, {Segransan}, {Sevin}, {Siebenmorgen}, {Soenke}, {Stadler}, {Suarez}, {Tiph{\`e}ne}, {Turatto}, {Udry}, {Vakili}, {Waters}, {Weber}, {Wildi}, {Zins}, \& {Zurlo}}]{Beuzit2019}
{Beuzit}, J.~L., {Vigan}, A., {Mouillet}, D., {et~al.} 2019, \aap, 631, A155

\bibitem[{{Biller} {et~al.}(2010){Biller}, {Liu}, {Wahhaj}, {Nielsen}, {Close}, {Dupuy}, {Hayward}, {Burrows}, {Chun}, {Ftaclas}, {Clarke}, {Hartung}, {Males}, {Reid}, {Shkolnik}, {Skemer}, {Tecza}, {Thatte}, {Alencar}, {Artymowicz}, {Boss}, {de Gouveia Dal Pino}, {Gregorio-Hetem}, {Ida}, {Kuchner}, {Lin}, \& {Toomey}}]{Biller2010}
{Biller}, B.~A., {Liu}, M.~C., {Wahhaj}, Z., {et~al.} 2010, \apjl, 720, L82

\bibitem[{{Biller} {et~al.}(2013){Biller}, {Liu}, {Wahhaj}, {Nielsen}, {Hayward}, {Males}, {Skemer}, {Close}, {Chun}, {Ftaclas}, {Clarke}, {Thatte}, {Shkolnik}, {Reid}, {Hartung}, {Boss}, {Lin}, {Alencar}, {de Gouveia Dal Pino}, {Gregorio-Hetem}, \& {Toomey}}]{Biller2013a}
{Biller}, B.~A., {Liu}, M.~C., {Wahhaj}, Z., {et~al.} 2013, \apj, 777, 160

\bibitem[{{Blunt} {et~al.}(2020){Blunt}, {Wang}, {Angelo}, {Ngo}, {Cody}, {De Rosa}, {Graham}, {Hirsch}, {Nagpal}, {Nielsen}, {Pearce}, {Rice}, \& {Tejada}}]{Blunt2020}
{Blunt}, S., {Wang}, J.~J., {Angelo}, I., {et~al.} 2020, \aj, 159, 89

\bibitem[{{Boss}(1997)}]{Boss1997}
{Boss}, A.~P. 1997, Science, 276, 1836

\bibitem[{{Bouvier} {et~al.}(2014){Bouvier}, {Matt}, {Mohanty}, {Scholz}, {Stassun}, \& {Zanni}}]{Bouvier2014}
{Bouvier}, J., {Matt}, S.~P., {Mohanty}, S., {et~al.} 2014, in Protostars and Planets VI, ed. H.~{Beuther}, R.~S. {Klessen}, C.~P. {Dullemond}, \& T.~{Henning}, 433

\bibitem[{{Bowler} {et~al.}(2020){Bowler}, {Blunt}, \& {Nielsen}}]{Bowler2020}
{Bowler}, B.~P., {Blunt}, S.~C., \& {Nielsen}, E.~L. 2020, \aj, 159, 63

\bibitem[{{Brandt}(2021)}]{Brandt2021}
{Brandt}, T.~D. 2021, \apjs, 254, 42

\bibitem[{{Chabrier} {et~al.}(2000){Chabrier}, {Baraffe}, {Allard}, \& {Hauschildt}}]{Chabrier2000}
{Chabrier}, G., {Baraffe}, I., {Allard}, F., \& {Hauschildt}, P. 2000, \apj, 542, 464

\bibitem[{{Chambers}(2021)}]{Chambers2021}
{Chambers}, J. 2021, \apj, 914, 102

\bibitem[{{Crepp} {et~al.}(2016){Crepp}, {Gonzales}, {Bechter}, {Montet}, {Johnson}, {Piskorz}, {Howard}, \& {Isaacson}}]{Crepp2016}
{Crepp}, J.~R., {Gonzales}, E.~J., {Bechter}, E.~B., {et~al.} 2016, \apj, 831, 136

\bibitem[{{Crepp} {et~al.}(2012){Crepp}, {Johnson}, {Fischer}, {Howard}, {Marcy}, {Wright}, {Isaacson}, {Boyajian}, {von Braun}, {Hillenbrand}, {Hinkley}, {Carpenter}, \& {Brewer}}]{Crepp2012}
{Crepp}, J.~R., {Johnson}, J.~A., {Fischer}, D.~A., {et~al.} 2012, \apj, 751, 97

\bibitem[{{de Regt} {et~al.}(2024){de Regt}, {Gandhi}, {Snellen}, {Zhang}, {Ginski}, {Gonz{\'a}lez Picos}, {Kesseli}, {Landman}, {Molli{\`e}re}, {Nasedkin}, {S{\'a}nchez-L{\'o}pez}, \& {Stolker}}]{deRegt2024}
{de Regt}, S., {Gandhi}, S., {Snellen}, I.~A.~G., {et~al.} 2024, \aap, 688, A116

\bibitem[{{Desidera} {et~al.}(2021){Desidera}, {Chauvin}, {Bonavita}, {Messina}, {LeCoroller}, {Schmidt}, {Gratton}, {Lazzoni}, {Meyer}, {Schlieder}, {Cheetham}, {Hagelberg}, {Bonnefoy}, {Feldt}, {Lagrange}, {Langlois}, {Vigan}, {Tan}, {Hambsch}, {Millward}, {Alcal{\'a}}, {Benatti}, {Brandner}, {Carson}, {Covino}, {Delorme}, {D'Orazi}, {Janson}, {Rigliaco}, {Beuzit}, {Biller}, {Boccaletti}, {Dominik}, {Cantalloube}, {Fontanive}, {Galicher}, {Henning}, {Lagadec}, {Ligi}, {Maire}, {Menard}, {Mesa}, {M{\"u}ller}, {Samland}, {Schmid}, {Sissa}, {Turatto}, {Udry}, {Zurlo}, {Asensio-Torres}, {Kopytova}, {Rickman}, {Abe}, {Antichi}, {Baruffolo}, {Baudoz}, {Baudrand}, {Blanchard}, {Bazzon}, {Buey}, {Carbillet}, {Carle}, {Charton}, {Cascone}, {Claudi}, {Costille}, {Deboulb{\'e}}, {De Caprio}, {Dohlen}, {Fantinel}, {Feautrier}, {Fusco}, {Gigan}, {Giro}, {Gisler}, {Gluck}, {Hubin}, {Hugot}, {Jaquet}, {Kasper}, {Madec}, {Magnard}, {Martinez}, {Maurel}, {Le Mignant}, {M{\"o}ller-Nilsson}, {Llored}, {Moulin}, {Orign{\'e}},
  {Pavlov}, {Perret}, {Petit}, {Pragt}, {Puget}, {Rabou}, {Ramos}, {Rigal}, {Rochat}, {Roelfsema}, {Rousset}, {Roux}, {Salasnich}, {Sauvage}, {Sevin}, {Soenke}, {Stadler}, {Suarez}, {Weber}, \& {Wildi}}]{Desidera2021}
{Desidera}, S., {Chauvin}, G., {Bonavita}, M., {et~al.} 2021, \aap, 651, A70

\bibitem[{{Dias} {et~al.}(2002){Dias}, {Alessi}, {Moitinho}, \& {L{\'e}pine}}]{Dias2002}
{Dias}, W.~S., {Alessi}, B.~S., {Moitinho}, A., \& {L{\'e}pine}, J.~R.~D. 2002, \aap, 389, 871

\bibitem[{{Dupuy} \& {Liu}(2017)}]{Dupuy2017}
{Dupuy}, T.~J. \& {Liu}, M.~C. 2017, \apjs, 231, 15

\bibitem[{{Fontanive} {et~al.}(2019){Fontanive}, {Mu{\v{z}}i{\'c}}, {}, {Bonavita}, \& {Biller}}]{Fontanive2019}
{Fontanive}, C., {Mu{\v{z}}i{\'c}}, {}, K., {Bonavita}, M., \& {Biller}, B. 2019, \mnras, 490, 1120

\bibitem[{{Forgan} {et~al.}(2015){Forgan}, {Parker}, \& {Rice}}]{Forgan2015}
{Forgan}, D., {Parker}, R.~J., \& {Rice}, K. 2015, \mnras, 447, 836

\bibitem[{{Forgan} {et~al.}(2018){Forgan}, {Hall}, {Meru}, \& {Rice}}]{Forgan2018}
{Forgan}, D.~H., {Hall}, C., {Meru}, F., \& {Rice}, W.~K.~M. 2018, \mnras, 474, 5036

\bibitem[{{Fujii} \& {Hori}(2019)}]{Fujii2019}
{Fujii}, M.~S. \& {Hori}, Y. 2019, \aap, 624, A110

\bibitem[{{Gagn{\'e}}(2024)}]{Gagne2024}
{Gagn{\'e}}, J. 2024, \pasp, 136, 063001

\bibitem[{{Gaia Collaboration} {et~al.}(2021){Gaia Collaboration}, {Brown}, {Vallenari}, {Prusti}, {de Bruijne}, {Babusiaux}, {Biermann}, {Creevey}, {Evans}, {Eyer}, {Hutton}, {Jansen}, {Jordi}, {Klioner}, {Lammers}, {Lindegren}, {Luri}, {Mignard}, {Panem}, {Pourbaix}, {Randich}, {Sartoretti}, {Soubiran}, {Walton}, {Arenou}, {Bailer-Jones}, {Bastian}, {Cropper}, {Drimmel}, {Katz}, {Lattanzi}, {van Leeuwen}, {Bakker}, {Cacciari}, {Casta{\~n}eda}, {De Angeli}, {Ducourant}, {Fabricius}, {Fouesneau}, {Fr{\'e}mat}, {Guerra}, {Guerrier}, {Guiraud}, {Jean-Antoine Piccolo}, {Masana}, {Messineo}, {Mowlavi}, {Nicolas}, {Nienartowicz}, {Pailler}, {Panuzzo}, {Riclet}, {Roux}, {Seabroke}, {Sordo}, {Tanga}, {Th{\'e}venin}, {Gracia-Abril}, {Portell}, {Teyssier}, {Altmann}, {Andrae}, {Bellas-Velidis}, {Benson}, {Berthier}, {Blomme}, {Brugaletta}, {Burgess}, {Busso}, {Carry}, {Cellino}, {Cheek}, {Clementini}, {Damerdji}, {Davidson}, {Delchambre}, {Dell'Oro}, {Fern{\'a}ndez-Hern{\'a}ndez}, {Galluccio}, {Garc{\'\i}a-Lario},
  {Garcia-Reinaldos}, {Gonz{\'a}lez-N{\'u}{\~n}ez}, {Gosset}, {Haigron}, {Halbwachs}, {Hambly}, {Harrison}, {Hatzidimitriou}, {Heiter}, {Hern{\'a}ndez}, {Hestroffer}, {Hodgkin}, {Holl}, {Jan{\ss}en}, {Jevardat de Fombelle}, {Jordan}, {Krone-Martins}, {Lanzafame}, {L{\"o}ffler}, {Lorca}, {Manteiga}, {Marchal}, {Marrese}, {Moitinho}, {Mora}, {Muinonen}, {Osborne}, {Pancino}, {Pauwels}, {Petit}, {Recio-Blanco}, {Richards}, {Riello}, {Rimoldini}, {Robin}, {Roegiers}, {Rybizki}, {Sarro}, {Siopis}, {Smith}, {Sozzetti}, {Ulla}, {Utrilla}, {van Leeuwen}, {van Reeven}, {Abbas}, {Abreu Aramburu}, {Accart}, {Aerts}, {Aguado}, {Ajaj}, {Altavilla}, {{\'A}lvarez}, {{\'A}lvarez Cid-Fuentes}, {Alves}, {Anderson}, {Anglada Varela}, {Antoja}, {Audard}, {Baines}, {Baker}, {Balaguer-N{\'u}{\~n}ez}, {Balbinot}, {Balog}, {Barache}, {Barbato}, {Barros}, {Barstow}, {Bartolom{\'e}}, {Bassilana}, {Bauchet}, {Baudesson-Stella}, {Becciani}, {Bellazzini}, {Bernet}, {Bertone}, {Bianchi}, {Blanco-Cuaresma}, {Boch}, {Bombrun}, {Bossini},
  {Bouquillon}, {Bragaglia}, {Bramante}, {Breedt}, {Bressan}, {Brouillet}, {Bucciarelli}, {Burlacu}, {Busonero}, {Butkevich}, {Buzzi}, {Caffau}, {Cancelliere}, {C{\'a}novas}, {Cantat-Gaudin}, {Carballo}, {Carlucci}, {Carnerero}, {Carrasco}, {Casamiquela}, {Castellani}, {Castro-Ginard}, {Castro Sampol}, {Chaoul}, {Charlot}, {Chemin}, {Chiavassa}, {Cioni}, {Comoretto}, {Cooper}, {Cornez}, {Cowell}, {Crifo}, {Crosta}, {Crowley}, {Dafonte}, {Dapergolas}, {David}, \& {David}}]{Gaia2021}
{Gaia Collaboration}, {Brown}, A.~G.~A., {Vallenari}, A., {et~al.} 2021, \aap, 649, A1

\bibitem[{{Galli} {et~al.}(2021){Galli}, {Bouy}, {Olivares}, {Miret-Roig}, {Sarro}, {Barrado}, \& {Berihuete}}]{Galli2021}
{Galli}, P.~A.~B., {Bouy}, H., {Olivares}, J., {et~al.} 2021, \aap, 654, A122

\bibitem[{{Gonz{\'a}lez Picos} {et~al.}(2025){Gonz{\'a}lez Picos}, {Snellen}, {de Regt}, {Landman}, {Zhang}, {Gandhi}, \& {S{\'a}nchez-L{\'o}pez}}]{Gonzalez2025}
{Gonz{\'a}lez Picos}, D., {Snellen}, I.~A.~G., {de Regt}, S., {et~al.} 2025, \aap, 693, A298

\bibitem[{{Gratton} {et~al.}(2024){Gratton}, {Bonavita}, {Mesa}, {Desidera}, {Zurlo}, {Marino}, {D'Orazi}, {Rigliaco}, {Nascimbeni}, {Barbato}, {Columba}, \& {Squicciarini}}]{Gratton2024}
{Gratton}, R., {Bonavita}, M., {Mesa}, D., {et~al.} 2024, \aap, 685, A119

\bibitem[{{Gratton} {et~al.}(2025){Gratton}, {Bonavita}, {Mesa}, {Desidera}, {Zurlo}, {Marino}, {D'Orazi}, {Rigliaco}, {Nascimbeni}, {Barbato}, {Columba}, \& {Squicciarini}}]{Gratton2025}
{Gratton}, R., {Bonavita}, M., {Mesa}, D., {et~al.} 2025, \aap, 694, A175

\bibitem[{{Hall} {et~al.}(2017){Hall}, {Forgan}, \& {Rice}}]{Hall2017}
{Hall}, C., {Forgan}, D., \& {Rice}, K. 2017, \mnras, 470, 2517

\bibitem[{Harris {et~al.}(2020)Harris, Millman, van~der Walt, Gommers, Virtanen, Cournapeau, Wieser, Taylor, Berg, Smith, Kern, Picus, Hoyer, van Kerkwijk, Brett, Haldane, del R{'{\i}}o, Wiebe, Peterson, G{'{e}}rard-Marchant, Sheppard, Reddy, Weckesser, Abbasi, Gohlke, \& Oliphant}]{Numpy2020}
Harris, C.~R., Millman, K.~J., van~der Walt, S.~J., {et~al.} 2020, Nature, 585, 357

\bibitem[{{H{\o}g} {et~al.}(2000){H{\o}g}, {Fabricius}, {Makarov}, {Urban}, {Corbin}, {Wycoff}, {Bastian}, {Schwekendiek}, \& {Wicenec}}]{Hog2000}
{H{\o}g}, E., {Fabricius}, C., {Makarov}, V.~V., {et~al.} 2000, \aap, 355, L27

\bibitem[{{Hunt} \& {Reffert}(2024)}]{Hunt2024}
{Hunt}, E.~L. \& {Reffert}, S. 2024, \aap, 686, A42

\bibitem[{Hunter(2007)}]{Matplotlib2007}
Hunter, J.~D. 2007, Computing In Science \& Engineering, 9, 90

\bibitem[{{Hwang} {et~al.}(2022){Hwang}, {Ting}, \& {Zakamska}}]{Hwang2022}
{Hwang}, H.-C., {Ting}, Y.-S., \& {Zakamska}, N.~L. 2022, \mnras, 512, 3383

\bibitem[{{Kerr} {et~al.}(2022){Kerr}, {Kraus}, {Murphy}, {Krolikowski}, {Bedding}, \& {Rizzuto}}]{Kerr2022}
{Kerr}, R., {Kraus}, A.~L., {Murphy}, S.~J., {et~al.} 2022, \apj, 941, 143

\bibitem[{{Konopacky} {et~al.}(2016){Konopacky}, {Marois}, {Macintosh}, {Galicher}, {Barman}, {Metchev}, \& {Zuckerman}}]{Konopacky2016}
{Konopacky}, Q.~M., {Marois}, C., {Macintosh}, B.~A., {et~al.} 2016, \aj, 152, 28

\bibitem[{{Krumholz} {et~al.}(2012){Krumholz}, {Klein}, \& {McKee}}]{Krumholz2012}
{Krumholz}, M.~R., {Klein}, R.~I., \& {McKee}, C.~F. 2012, \apj, 754, 71

\bibitem[{{Maire} {et~al.}(2020){Maire}, {Baudino}, {Desidera}, {Messina}, {Brandner}, {Godoy}, {Cantalloube}, {Galicher}, {Bonnefoy}, {Hagelberg}, {Olofsson}, {Absil}, {Chauvin}, {Henning}, \& {Langlois}}]{Maire2020}
{Maire}, A.~L., {Baudino}, J.~L., {Desidera}, S., {et~al.} 2020, \aap, 633, L2

\bibitem[{{Maire} {et~al.}(2021){Maire}, {Langlois}, {Delorme}, {Chauvin}, {Gratton}, {Vigan}, {Girard}, {Wahhaj}, {Pott}, {Burtscher}, {Boccaletti}, {Carlotti}, {Henning}, {Kenworthy}, {Kervella}, {Rickman}, \& {Schmidt}}]{Maire2021}
{Maire}, A.-L., {Langlois}, M., {Delorme}, P., {et~al.} 2021, Journal of Astronomical Telescopes, Instruments, and Systems, 7, 035004

\bibitem[{{Mamajek}(2016)}]{Mamajek2016}
{Mamajek}, E.~E. 2016, in IAU Symposium, Vol. 314, Young Stars \& Planets Near the Sun, ed. J.~H. {Kastner}, B.~{Stelzer}, \& S.~A. {Metchev}, 21--26

\bibitem[{{Marois} {et~al.}(2006){Marois}, {Lafreni{\`e}re}, {Doyon}, {Macintosh}, \& {Nadeau}}]{Marois2006}
{Marois}, C., {Lafreni{\`e}re}, D., {Doyon}, R., {Macintosh}, B., \& {Nadeau}, D. 2006, \apj, 641, 556

\bibitem[{{Nielsen} {et~al.}(2019){Nielsen}, {De Rosa}, {Macintosh}, {Wang}, {Ruffio}, {Chiang}, {Marley}, {Saumon}, {Savransky}, {Ammons}, {Bailey}, {Barman}, {Blain}, {Bulger}, {Burrows}, {Chilcote}, {Cotten}, {Czekala}, {Doyon}, {Duch{\^e}ne}, {Esposito}, {Fabrycky}, {Fitzgerald}, {Follette}, {Fortney}, {Gerard}, {Goodsell}, {Graham}, {Greenbaum}, {Hibon}, {Hinkley}, {Hirsch}, {Hom}, {Hung}, {Dawson}, {Ingraham}, {Kalas}, {Konopacky}, {Larkin}, {Lee}, {Lin}, {Maire}, {Marchis}, {Marois}, {Metchev}, {Millar-Blanchaer}, {Morzinski}, {Oppenheimer}, {Palmer}, {Patience}, {Perrin}, {Poyneer}, {Pueyo}, {Rafikov}, {Rajan}, {Rameau}, {Rantakyr{\"o}}, {Ren}, {Schneider}, {Sivaramakrishnan}, {Song}, {Soummer}, {Tallis}, {Thomas}, {Ward-Duong}, \& {Wolff}}]{Nielsen2019}
{Nielsen}, E.~L., {De Rosa}, R.~J., {Macintosh}, B., {et~al.} 2019, \aj, 158, 13

\bibitem[{{Nielsen} {et~al.}(2012){Nielsen}, {Liu}, {Wahhaj}, {Biller}, {Hayward}, {Boss}, {Bowler}, {Kraus}, {Shkolnik}, {Tecza}, {Chun}, {Clarke}, {Close}, {Ftaclas}, {Hartung}, {Males}, {Reid}, {Skemer}, {Alencar}, {Burrows}, {de Gouveia Dal Pino}, {Gregorio-Hetem}, {Kuchner}, {Thatte}, \& {Toomey}}]{Nielsen2012}
{Nielsen}, E.~L., {Liu}, M.~C., {Wahhaj}, Z., {et~al.} 2012, \apj, 750, 53

\bibitem[{{{\"O}berg} {et~al.}(2011){{\"O}berg}, {Murray-Clay}, \& {Bergin}}]{Oberg2011}
{{\"O}berg}, K.~I., {Murray-Clay}, R., \& {Bergin}, E.~A. 2011, \apjl, 743, L16

\bibitem[{{Parker} \& {Quanz}(2012)}]{Parker2012}
{Parker}, R.~J. \& {Quanz}, S.~P. 2012, \mnras, 419, 2448

\bibitem[{{Peretti} {et~al.}(2019){Peretti}, {S{\'e}gransan}, {Lavie}, {Desidera}, {Maire}, {D'Orazi}, {Vigan}, {Baudino}, {Cheetham}, {Janson}, {Chauvin}, {Hagelberg}, {Menard}, {Heng}, {Udry}, {Boccaletti}, {Daemgen}, {Le Coroller}, {Mesa}, {Rouan}, {Samland}, {Schmidt}, {Zurlo}, {Bonnefoy}, {Feldt}, {Gratton}, {Lagrange}, {Langlois}, {Meyer}, {Carbillet}, {Carle}, {De Caprio}, {Gluck}, {Hugot}, {Magnard}, {Moulin}, {Pavlov}, {Pragt}, {Rabou}, {Ramos}, {Rousset}, {Sevin}, {Soenke}, {Stadler}, {Weber}, \& {Wildi}}]{peretti2019}
{Peretti}, S., {S{\'e}gransan}, D., {Lavie}, B., {et~al.} 2019, \aap, 631, A107

\bibitem[{{Pollack} {et~al.}(1996){Pollack}, {Hubickyj}, {Bodenheimer}, {Lissauer}, {Podolak}, \& {Greenzweig}}]{Pollack1996}
{Pollack}, J.~B., {Hubickyj}, O., {Bodenheimer}, P., {et~al.} 1996, \icarus, 124, 62

\bibitem[{{Savvidou} \& {Bitsch}(2023)}]{Savvidou2023}
{Savvidou}, S. \& {Bitsch}, B. 2023, \aap, 679, A42

\bibitem[{{Schwarz} {et~al.}(2016){Schwarz}, {Ginski}, {de Kok}, {Snellen}, {Brogi}, \& {Birkby}}]{Schwarz2016}
{Schwarz}, H., {Ginski}, C., {de Kok}, R.~J., {et~al.} 2016, \aap, 593, A74

\bibitem[{{Shkolnik} {et~al.}(2017){Shkolnik}, {Allers}, {Kraus}, {Liu}, \& {Flagg}}]{Shkolnik2017}
{Shkolnik}, E.~L., {Allers}, K.~N., {Kraus}, A.~L., {Liu}, M.~C., \& {Flagg}, L. 2017, \aj, 154, 69

\bibitem[{{Stamatellos} \& {Whitworth}(2009)}]{Stamatellos2009}
{Stamatellos}, D. \& {Whitworth}, A.~P. 2009, \mnras, 392, 413

\bibitem[{{Stolker} {et~al.}(2020){Stolker}, {Quanz}, {Todorov}, {K{\"u}hn}, {Molli{\`e}re}, {Meyer}, {Currie}, {Daemgen}, \& {Lavie}}]{Stolker2020b}
{Stolker}, T., {Quanz}, S.~P., {Todorov}, K.~O., {et~al.} 2020, \aap, 635, A182

\bibitem[{{van Elteren} {et~al.}(2019){van Elteren}, {Portegies Zwart}, {Pelupessy}, {Cai}, \& {McMillan}}]{vanElteren2019}
{van Elteren}, A., {Portegies Zwart}, S., {Pelupessy}, I., {Cai}, M.~X., \& {McMillan}, S.~L.~W. 2019, \aap, 624, A120

\bibitem[{{Vigan}(2020)}]{Vigan2020}
{Vigan}, A. 2020, {vlt-sphere: Automatic VLT/SPHERE data reduction and analysis}

\bibitem[{{Wahhaj} {et~al.}(2011){Wahhaj}, {Liu}, {Biller}, {Clarke}, {Nielsen}, {Close}, {Hayward}, {Mamajek}, {Cushing}, {Dupuy}, {Tecza}, {Thatte}, {Chun}, {Ftaclas}, {Hartung}, {Reid}, {Shkolnik}, {Alencar}, {Artymowicz}, {Boss}, {de Gouveia Dal Pino}, {Gregorio-Hetem}, {Ida}, {Kuchner}, {Lin}, \& {Toomey}}]{Wahhaj2011}
{Wahhaj}, Z., {Liu}, M.~C., {Biller}, B.~A., {et~al.} 2011, \apj, 729, 139

\bibitem[{{Zhang} {et~al.}(2021{\natexlab{a}}){Zhang}, {Snellen}, {Bohn}, {Molli{\`e}re}, {Ginski}, {Hoeijmakers}, {Kenworthy}, {Mamajek}, {Meshkat}, {Reggiani}, \& {Snik}}]{Zhangyp2021a}
{Zhang}, Y., {Snellen}, I. A.~G., {Bohn}, A.~J., {et~al.} 2021{\natexlab{a}}, \nat, 595, 370

\bibitem[{{Zhang} {et~al.}(2021{\natexlab{b}}){Zhang}, {Snellen}, \& {Molli{\`e}re}}]{Zhangyp2021b}
{Zhang}, Y., {Snellen}, I. A.~G., \& {Molli{\`e}re}, P. 2021{\natexlab{b}}, \aap, 656, A76

\bibitem[{{Zheng} {et~al.}(2015){Zheng}, {Kouwenhoven}, \& {Wang}}]{Zheng2015}
{Zheng}, X., {Kouwenhoven}, M.~B.~N., \& {Wang}, L. 2015, \mnras, 453, 2759

\bibitem[{{Zuckerman}(2019)}]{Zuckerman2019a}
{Zuckerman}, B. 2019, \apj, 870, 27

\bibitem[{{Zuckerman} {et~al.}(2019){Zuckerman}, {Klein}, \& {Kastner}}]{Zuckerman2019}
{Zuckerman}, B., {Klein}, B., \& {Kastner}, J. 2019, \apj, 887, 87

\end{thebibliography}

% %%%%%%%%%%%%%%%%%%%%%%%%%%%%%%%%%%%%%%%%%%%%%%%%%%%%%%%%%%%%%%
% Example below of non-structurated natbib references
% To use the v8.3 macros with this form of composition of bibliography,
% the option "bibyear" should be added to the command line
% "\documentclass[bibyear]{aa}".
% %%%%%%%%%%%%%%%%%%%%%%%%%%%%%%%%%%%%%%%%%%%%%%%%%%%%%%%%%%%%%%

% \begin{thebibliography}{}

%   \bibitem[1966]{baker} Baker, N. 1966,
%       in Stellar Evolution,
%       ed.\ R. F. Stein,\& A. G. W. Cameron
%       (Plenum, New York) 333

%    \bibitem[1988]{balluch} Balluch, M. 1988,
%       A\&A, 200, 58

%    \bibitem[1980]{cox} Cox, J. P. 1980,
%       Theory of Stellar Pulsation
%       (Princeton University Press, Princeton) 165

%    \bibitem[1969]{cox69} Cox, A. N.,\& Stewart, J. N. 1969,
%       Academia Nauk, Scientific Information 15, 1

%    \bibitem[1980]{mizuno} Mizuno H. 1980,
%       Prog. Theor. Phys., 64, 544

%    \bibitem[1987]{tscharnuter} Tscharnuter W. M. 1987,
%       A\&A, 188, 55

%    \bibitem[1992]{terlevich} Terlevich, R. 1992, in ASP Conf. Ser. 31,
%       Relationships between Active Galactic Nuclei and Starburst Galaxies,
%       ed. A. V. Filippenko, 13

%    \bibitem[1980a]{yorke80a} Yorke, H. W. 1980a,
%       A\&A, 86, 286

%    \bibitem[1997]{zheng} Zheng, W., Davidsen, A. F., Tytler, D. \& Kriss, G. A.
%       1997, preprint
% \end{thebibliography}

%%%%%%%%%%%%%%%%%%%%%%%%%%%%%%%%%%%%%%%%%%%%%%%%%%%%%%%%%%%%%%%
% Appendices must be placed after   \end{thebibliography}
% They will be placed automatically on a new page.
%%%%%%%%%%%%%%%%%%%%%%%%%%%%%%%%%%%%%%%%%%%%%%%%%%%%%%%%%%%%%%%
\begin{appendix}
%%%%%%%%%%%%%%%%%%%%%%%%%%%%%%%%%%%%%%%%%%%%%%%%%%%%%%%%%%%%%%%
% In the PDF output, floats should be placed
% under their own appendix, not before the title, nor after the
% title of the next appendix.

% In short appendices, onecolumn floats (\figure*
% or \table*) will generate a blank page.
% To prevent this behaviour, a few examples are provided here.

% In case you have a lot of floating objects for little text and the
% LaTeX engine moves the floats away from their context, the command
% \FloatBarrier of the “placeins” package will empty the
% float buffer and place all stored floats in the continuity.

% If you still encounter problems with wide floats placement,
% just use the onecolumn environment throughout the appendices.
%%%%%%%%%%%%%%%%%%%%%%%%%%%%%%%%%%%%%%%%%%%%%%%%%%%%%%%%%%%%%%%

%____________________________________________________________
%       Wide floats at the start of an appendix: first method
%-------------------------------------------------------------
% To prevent a blank page after the start of an appendix:
% - Switch to one \onecolumn first
% - Declare the section title
% - Declare the onecolumn float with the parameter [h!]
% - Revert to \twocolumn at the end of the section
\twocolumn
\section{SED fitting}
\label{app:sed}
\begin{figure}[h!]
    \centering
     \includegraphics[width=0.5\textwidth]{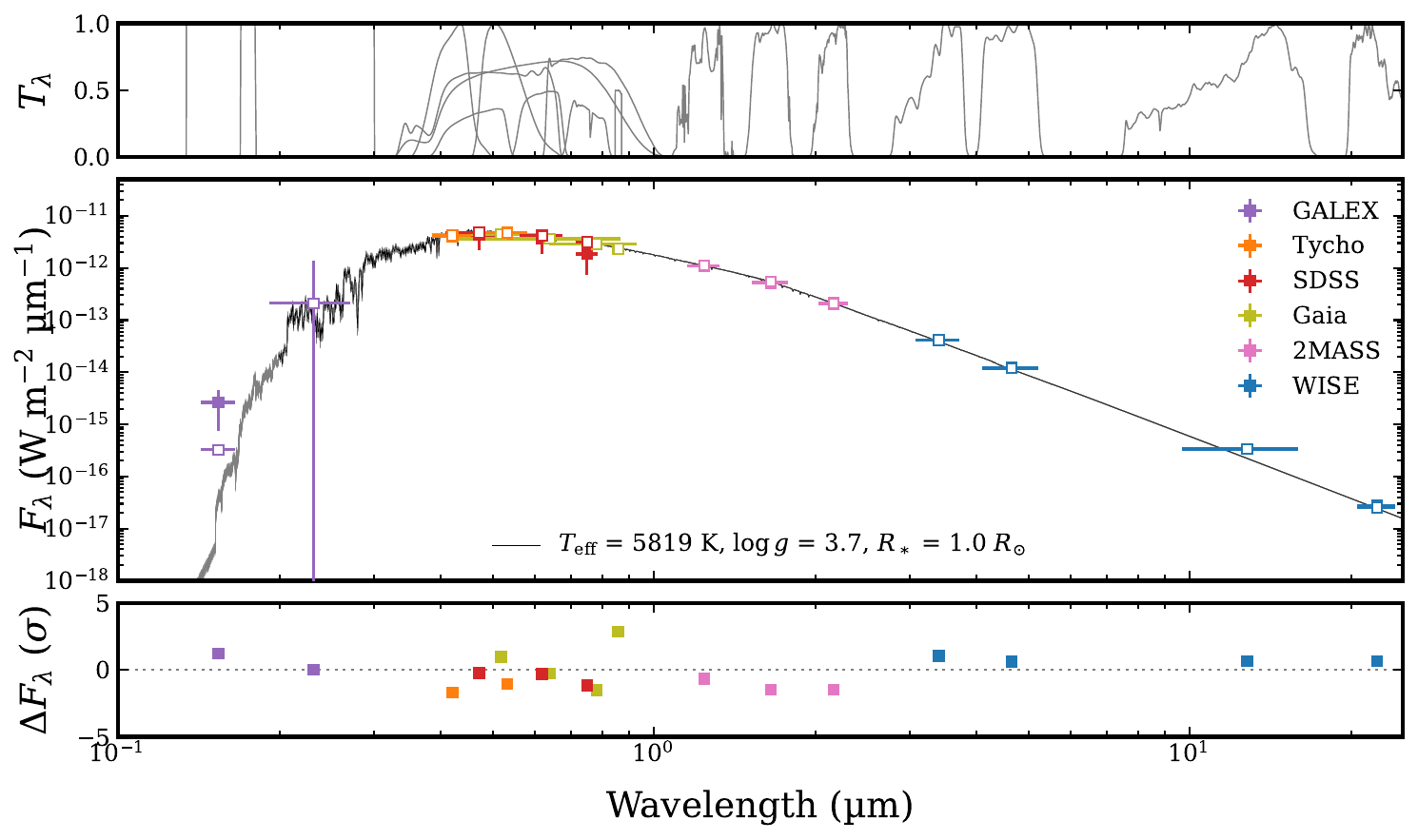}
     \includegraphics[width=0.5\textwidth]{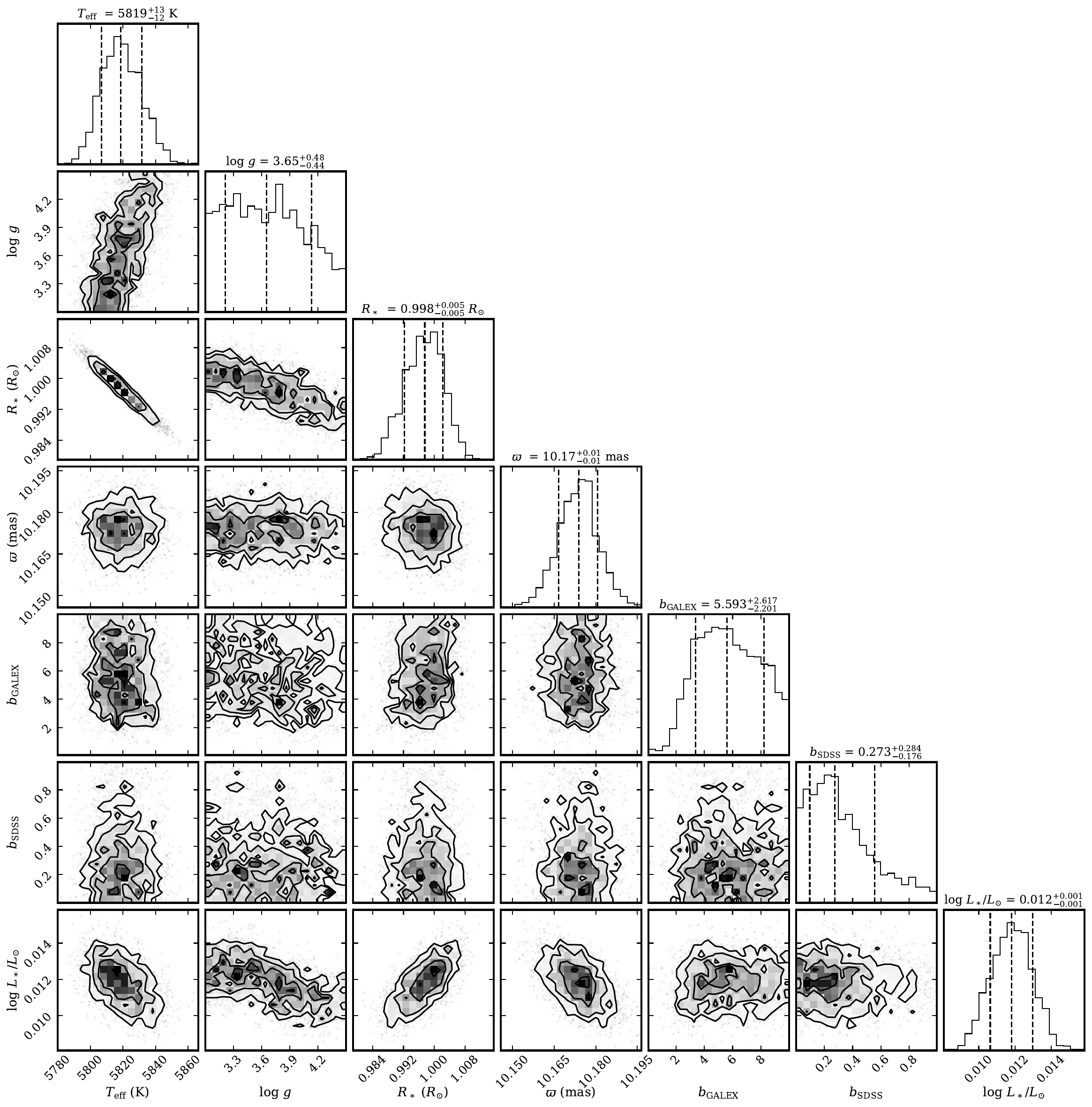}
     \caption{SED fitting of the central star using BT-Settl-CIFIST models. \rev{Upper panel: the filled squares with errorbar are the measured photometric data and the open squares are the modelled photometric data. The black line is the optimised model spectrum and the grey lines are 30 spectra randomly drawn from the posterior distributions.
     Lower panel: posteriors distributions of SED fitting.}
     $b_{GALEX}$ and $b_{SDSS}$ are the inflation factors of the GALEX and SDSS photometric data.}
    \label{fig:star_SED}
\end{figure}

We present the posterior distribution of the SED fitting of the primary star in Fig.~\ref{fig:star_SED}. We included error inflation into fitting because the uncertainty in GALEX and SDSS photometric data is likely underestimated.
No infrared excess is detected.
% The fitted mass here is not well-constrained as it is related to the surface gravity which is not well-constrained in the SED fitting. We adopted the mass of 1.06\,\msun from \cite{Kerr2022}.

\begin{figure}[h!]
    \centering
     \includegraphics[width=0.5\textwidth]{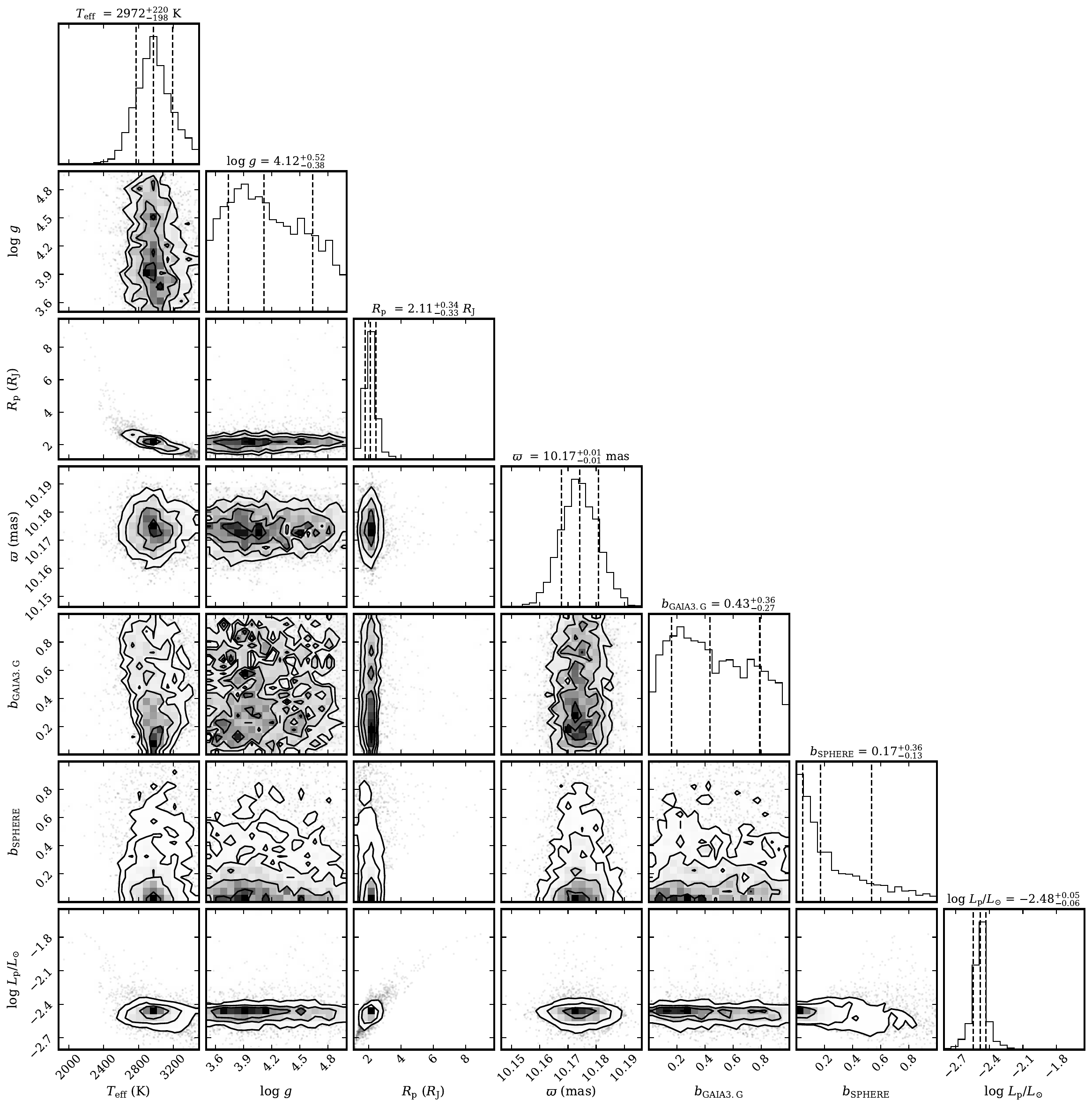}
     \caption{Posterior distribution of the SED fitting of the companion using BT-Settl models. $b_{GAIA.G}$ and $b_{SPHERE}$ are the inflation factors of the Gaia and SPHERE photometric data}
    \label{fig:comsed_post}
\end{figure}

The posterior distribution of the SED fitting of the companion is shown in Fig.~\ref{fig:comsed_post}.
% The fitted mass here is not meaningful here either because it is related to the surface gravity which is not well-constrained.

\section{Orbital fitting}
\begin{figure}[h!]
    \centering
     \includegraphics[width=0.5\textwidth]{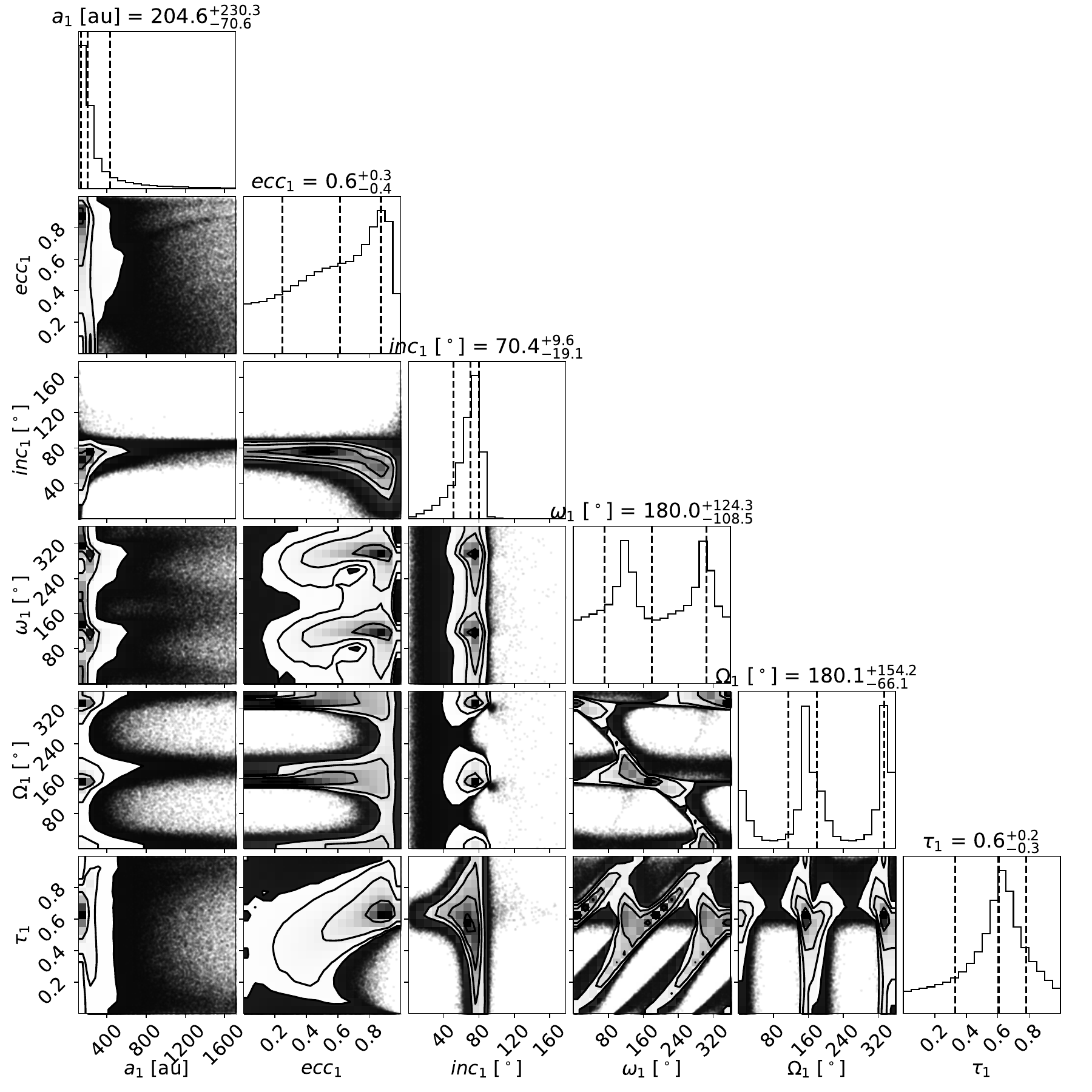}
     \caption{}
     \label{fig:orbits_post}
\end{figure}

We show the posterior distribution of orbital parameters of H~24121B in Fig.~\ref{fig:orbits_post}.

\label{app:vel}

\begin{figure}[h!]
    \centering
     \includegraphics[width=0.5\textwidth]{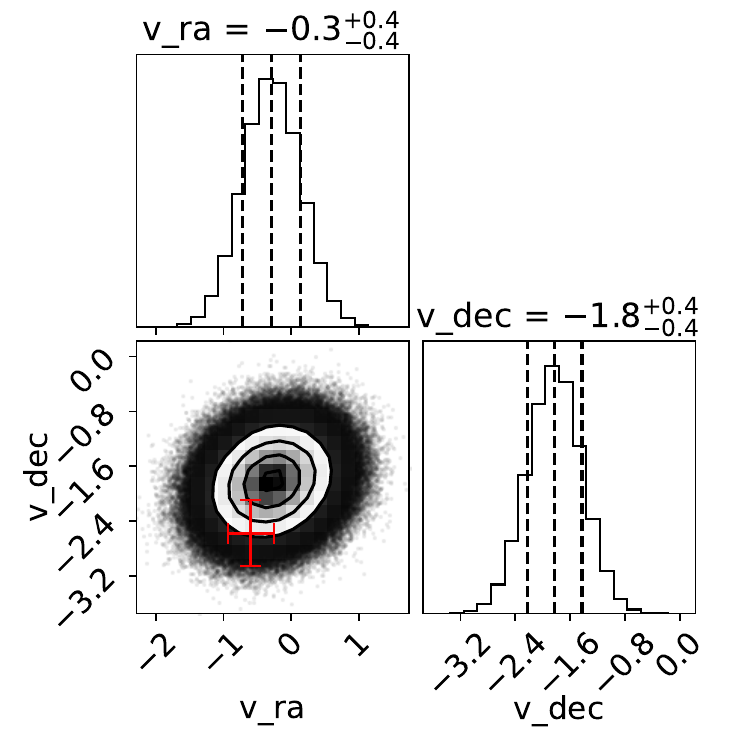}
     \caption{Posterior distribution of the orbital velocity of HD~24121B in the Gaia reference epoch. The units are in mas yr$^{-1}$. The red data point is the proper motion difference from Gaia.}
    \label{fig:vel_post}
\end{figure}

We show the posterior distribution of orbital velocity from \textsc{orbitize} fitting in the Gaia reference epoch 2016.0 in Fig.~\ref{fig:vel_post}. The orbital velocity agrees with the proper motion difference measured by Gaia within 1--2$\sigma$.

% If text follows like so
% it is easier to finish the page in onecolumn and revert to
% twocolumn when starting the next page if needed.}

% \FloatBarrier %\usepackage{placeins}
% \twocolumn
%____________________________________________________________
%       Wide floats at the start of an appendix: second method
%-------------------------------------------------------------
% To prevent a blank page, a second method is:
% - Declare the onecolumn float *
% - Declare the section under the float
% However, this method should be reserved to appendices
% containing only onecolumn tables or figures.

% To prevent a blank page, a second method is to insert
% the appendix title \underline{after} declaring the onecolumn float.
% \newline This method should be reserved to appendices
% containing only one-column floats\{figure*\} or \{table*\}
% and no text.

\end{appendix}
\end{document}